\documentclass[11pt,aps]{revtex4}
\usepackage{epsfig}
\usepackage{amsmath,amssymb,color}
\usepackage{epstopdf}
\parskip=\medskipamount

\newcommand{\beq}{\begin{equation}}
\newcommand{\eeq}{\end{equation}}
\newcommand{\be}{\begin{equation}}
\newcommand{\ee}{\end{equation}}

\def \ll {\langle}

\begin{document}

\title{Mobility edge and Black Hole Horizon}

\author{A. Gorsky$^{1,2}$}

\affiliation{
$^1$ Institute of Information Transmission Problems of the Russian Academy of Sciences, Moscow,
Russia, \\
$^2$ Moscow Institute of Physics and Technology, Dolgoprudny 141700, Russia 
}

\begin{abstract}
We conjecture that the mobility edge in the 4D Euclidean Dirac operator
spectrum in QCD in the deconfined phase found in the lattice studies 
corresponds to the
near black hole (BH) horizon region in the holographic dual. We present
some evidences both from the field theory side and from the
worldsheet theory of long open string.  

\end{abstract}
\maketitle
\section{Introduction} 
In this talk I shall mainly follow the 
conjectures  and questions formulated in \cite{gls}. First, let me  explain the motivation
of our study. The 4D Euclidean Dirac operator spectrum in QCD is the important
observable both in the confined and deconfined phases.
The famous Casher-Banks relation \cite{casher}  for the chiral condensate 
in terms of the spectral density at the origin in the confined phase is 
the most immediate example. The interesting
sum rules can be formulated for  eigenvalues \cite{sumrules} and
two types of  matrix models provide the powerful tools for a
investigation of the Dirac operator spectral properties (see \cite{matrix} for review).
One matrix model corresponds to the zero momentum sector of the Chiral Lagrangian while
the second model mimics the fermion determinant integrated over the moduli space of a  
instanton-antiinstanton ensemble in QCD ground state.

It was suggested in \cite{stern} to attack the spectral problem for the
Dirac operator by the tools familiar in solid state physics and treat the Euclidean 
Dirac operator in 4D as the Hamiltonian with respect to the additional
fifth time coordinate identified with the Schwinger proper time. This viewpoint allowed
to identify the pion as the analogue of the diffuson,  relate the 
diffusion constant with the pion decay constant and identify the 
standard energy and time scales familiar in the disordered systems.

According to this logic the question concerning the localization of the eigenmodes 
of the Dirac operator is  quite natural.
The spectral formfactor and spectral correlator are standard and quite informative
objects which allow to identify the localization properties of the
eigenfunctions. The more involved quantity is the
level spacing distribution which distinguishes the  parts of the spectrum
with  localized and delocalized modes. The delocalized
modes are subject of the level interaction and obey the Wigner-Dyson
statistics while  the localized modes do not correlate and obey the
Poisson statistics. There can be the mobility edge separating the 
localized and delocalized modes in the $d\geq 3$ where d is dimension
of space. 

The localization properties of the Dirac operator spectrum in QCD 
have been investigated in the lattice version and it has been found
a bit surprisingly that all modes are delocalized in the confined phase while 
there is the mobility edge in the deconfined phase  \cite{osborn}. Low energy
modes in the deconfined phase are localized while high energy part of the
spectrum is delocalized. Moreover it has been found \cite{kovacs} that the
mobility edge appears exactly at the temperature of the deconfinement phase transition.

In \cite{gls} we attack the issue of the Dirac operator spectrum from the
holographic viewpoint. The first indication of the holographic
picture behind the scene is the relation between the Schwinger proper time
and the radial coordinate in the ADS space \cite{gopakumar}. This relation
still holds for the case of the external field \cite{gorly} and  finite
temperature \cite{furuuchi}. On the other hand the dynamics in the 
radial direction amounts the holographic RG flows \cite{dbvv}.  
Hence the time conjugated to the 4D Dirac operator considered 
as the Hamiltonian in 4D Euclidean space is the RG
time in holographic RG flows and many questions can be 
reformulated in RG terms.

The deconfinement transition in holographic
QCD corresponds to the change of the bulk geometry which involves
the thermal AdS at $T<T_c$ and AdS BH at $T>T_c$ \cite{witten98,aharonydec}. Hence it is natural 
to question how the emergence of BH in the deconfined phase and 
the mobility edge in the 4D Dirac operator spectrum are correlated. We
have found the evidences that indeed the corresponding energy scales - radius of BH
and the mobility edge are approximately the same. We shall use the
field theory arguments as well
as the arguments based on the worldsheet theory of a long open string extended
along the radial coordinate in AdS which represents the matter in 
fundamental representation.

Let us emphasize from the very beginning that the localization phase transition  
in the Dirac operator spectrum in QCD which we discuss in this study 
is not  the conventional Anderson-type transition. 
Indeed the phase transitions driven by disorder depend on the
dimensionality and the dispersion relation for the degrees of freedom. It was observed long time ago
in \cite{fradkin} that the disorder driven transition for the
Dirac operator differs from the transition for the Schrodinger
operator.  The key difference
concerns the role of spectral density  as the order parameter.
In the Anderson transition  the spectral density does not play
any essential role and only a spectral formfactor 
and higher spectral correlators matter. On the
other hand for the disordered Dirac operator in higher dimensions
the spectral density at the origin becomes the order parameter.
Moreover it was argued recently \cite{radzi} that the non-Anderson disorder
driven phase transition takes place if
\beq
d>2\gamma, \qquad  E\propto k^{\gamma}
\eeq
where $\gamma$ defines the dispersion relation, $\gamma=1$ for the 
Dirac operator. The non-Anderson transition can be supplemented
with the additional localization transition if it is allowed 
by the symmetry.

We shall be interested in the Dirac operator spectrum in $4d$
Euclidean space hence indeed the non-Anderson transition can be expected. The
phase transition with the chiral symmetry restoration results at $T=T_c$ to  vanishing of
spectral density at the origin while at higher temperatures
in the deconfined phase the mobility edge indicating the 
localization transition appears.

\section{Dirac operator in QCD}
Let us recall some results concerning the Dirac operator spectrum.
The partition function of QCD reads as 
\beq
Z_{QCD}= \int dA_{\mu}\prod_{f=1}^{N_f} det(iD+m_f)\exp(-S_{YM}(A))
\eeq
and can be decomposed into the sectors with fixed topological charges weighted
with the $\theta$ term
\beq
Z_{QCD}(\theta)=\sum_{\nu} e^{i\theta\nu} Z_{\nu,QCD}
\eeq
We shall deal with the eigenvalue equation  for the 4D Euclidean Dirac operator
\beq
\hat{D}(A)\psi_n=i\lambda_n\psi_n
\eeq
which coincides with the Dirac equation for the imaginary fermion 
mass $m=i\lambda$.
The spectral density is defined as 
\beq
\rho(\lambda)=<\sum_n \delta(\lambda- \lambda_n)>_{QCD}
\eeq
which at the origin in the confined phase yields the chiral condensate \cite{casher}
\beq
<\bar{\Psi}\Psi>= \Sigma =\frac{\pi \rho(0)}{V}
\eeq
The spectral density can be derived from the discontinuity of the resolvent
across the imaginary axis
\beq
\Sigma(z)=<V^{-1}Tr\frac{1}{D+z}>
\eeq

Let us summarize the results concerning the spectral correlators for Dirac operator
in the confining phase \cite{matrix}.
In the bulk of the spectrum where $\lambda\propto O(1)$ at large N one has
\beq
R(x_1,x_2) = det(K(x_i,x_j))_{i,j=1,2}
\eeq
\beq
K(x-y)=\frac{sin\pi(x-y)}{\pi (x-y)}
\eeq
In the microscopic region near the origin the kernel gets modified to 
\beq
K(x-y) = \sqrt{xy} \frac{xJ_{\alpha +1}(x)J_{\alpha}(y)  - yJ_{\alpha}(x)J_{\alpha+1}(y)}{x^2 -y^2}
\eeq
where $J_{\alpha}$ denotes Bessel function and $\alpha= N_f +|\nu|$. The microscopic
kernel is sensitive to the topological sector and the microscopic spectral density
is related to the kernel $\rho_s(z)= K(z,z)$
\beq
\rho_s(z)= \frac{z}{2}[J^2_{N_f+|\nu|}(z) - J_{N_f+|\nu|+1 }(z)J_{N_f+|\nu|-1}(z)]
\eeq
This spectral density can be obtained from the partially quenched Chiral Lagrangian.
The kernel in the microscopic region can be expressed in terms of the partition function
with two additional flavors of imaginary masses $m_1,m_2$ \cite{damgaard}
\beq
K(x,y)= \frac{1}{2}(xy)^{N_f+1/2}\frac{Z_{N_f+2}(m_1=ix, m_2=iy)}{Z_{N_f}}
\eeq

The lattice studies definitely demonstrate \cite{osborn,kovacs} that 
all eigenvalues of 4d Euclidean Dirac operator in the confined phase  are delocalized
hence it behaves as 4d metal.
The metallic property of the Dirac operator in confined phase of QCD is quite counterintuitive
since quark is assumed to be confined in 3d space. The most naive 
argument in favor of the metal phase is that  at $T=0$ we have $O(4)$ symmetry
in Euclidean space and expected delocalization in Euclidean time implies
the delocalization in the 3d space. However since Polyakov line vanishes 
in the confined phase there is no $O(4)$ symmetry breaking
disorder.  Another standard argument applied to the zero eigenvalue only is that
the fermionic zero modes get delocalized in the instanton-antiinstanton ensemble.
Argument of Parisi \cite{parisi} claims that 
the eigenvalues of the Dirac operator in confined phase  have to interact to provide 
the finite density at the origin and hence obey the Wigner-Dyson statistics.

Here we suggest that  the topological delocalization phenomenon 
namely that the disorder driven  localization transition 
is strongly influenced by the topological terms \cite{furusaki,kamenev1}
is relevant.
If the topological terms are added to the action
the two-couplings RG flow occurs and some universal 
phenomena get manifested. Namely at some quantized values 
of the chemical potentials for the topological terms the 
delocalized modes appear in the spectrum. There is example of
$"'\theta =\pi"'$ phenomena   in the QHE effect \cite{khmel, pruisken}.
The peculiar aspects of such regime have been
recently reconsidered in 4D Yang-Mills theory \cite{seiberg}. Generically there is no 
mobility edge in disordered (1+1) and (2+1) systems however 
the topological delocalization phenomenon occurs at some
values of the chemical potential for the winding number.
It is the manifestation of the instanton effects
in the two-couplings RG flows.

We conjecture that the similar topological delocalization phenomenon occurs 
in the $(4+1)$ case as well in the confined phase.
The mass dependence  due to the axial anomaly goes through $m^{N_f}e^{i\theta}$ 
hence to have imaginary masses as required for eigenvalues of the Dirac operator we 
can introduce the particular value of $\theta$ which depends on the number
of flavors. In particular for $N_f=0$ when the Dirac operator plays 
the role of probe $\theta=\pi$ and generically $\theta=\frac{\pi (N_f+2)}{2}$. 
With this interpretation the delocalization of Dirac operator 
modes occurs only at the particularly values of $\theta$. 

\section{Aspects of Holographic  QCD}

   Turn to the holographic QCD and consider the Witten-Sakai-Sugimoto geometry \cite{witten98,ss}
It involves the $N_c$ D4 branes wrapped around the cylinder with the boundary
conditions breaking SUSY. At large $N_s$ the D4 branes pinch the cylinder which
turns into the cigar. The total 10d geometry
looks as $R^{3,1}\times S^4 \times (r,\phi)$ at small temperature.
Let drop off all space-time and $S^4$ dependence of the fields
we are interested in for a moment. Hence the relevant part
of the holographic geometry is the cigar in $(r,\phi)$ coordinates
\beq
ds^2=(r/R)^{3/2}f(r)d\phi^2 + (R/r)^{3/2}\frac{dr^2}{f(r)} \qquad f(r)= 1 -(\frac{r_{kk}}{r})^3 
\eeq 
where $\phi$ is periodic variable.
Metric at low temperature is close to the thermal $AdS_2$ . We insert D8-branes extended 
in radial coordinate r and localized at $\phi$ and D0 branes 
localized in radial coordinate in this background geometry
and extended along $\phi$. 
The $N_f$  $D8-\bar{D8}$ 
branes are connected at the tip of the cigar and are placed at 
$\phi=0,\pi$ on the $\phi$ circle.
The D8 branes carries $U(N_F)$ flavor gauge group
at the worldvolume and matrix U of the pseudoscalar mesons $\pi_a$ is defined in terms of 
holonomy of radial component of the flavor
gauge field 
\beq
U= e^{it_a\pi_a}= e^{\int A_rdr}
\eeq
Above the critical temperature the metric is identified with the AdS BH in $N=4$ theory \cite{witten98}
and the phase transition in QCD qualitatively corresponds to the Hawking-Page transition.
At $T=T_c$ the $\phi$ and $t_E$ 
coordinates get interchanged.

It is worth also to comment on the origin of the mass term $TrMU$  in the Chiral Lagrangian.
It was argued in \cite{aharony2,hashimoto} that it  comes from the 
worldsheet instanton that is the open string with worldsheet $(r,\phi)$ coordinates
which is stretched between left and right D8 branes and spans some area on the cigar.
The mass comes from the Nambu-Goto string action while the factor U comes 
from the interaction of the open string end with D8 brane. We are interested 
in the Dirac operator eigenvalues that is purely imaginary masses. They can be
obtained if the $\theta$ term is taken into account which holographically corresponds
to the holonomy of the RR 1-form field along KK circle \cite{wittentheta}. With the proper
value of $\theta$ we get purely imaginary masses.

Remark that  random fields act on open string not only at its ends. Indeed the QCD vacuum
is populated by instantons and antiinstantons which are extended in $\phi$.
The  position of D0 instanton
along the radial coordinate is known to be related with its size $\sigma$. Therefore
density  $d(\sigma)\propto \sigma^{-5}$ gets mapped into
the density of the instantons  along the radial coordinate.The instantons and
antiinstantons act as random forces acting on the string worldsheet. 
The density of the external forces is r-dependent and 
the strength of the force follows from the "`instanton mass"- its action
$\exp(-\frac{8\pi}{g^2(r)} +i \theta)$.

\section{Critical regime}
\subsection{Diagnostics of the critical behavior}

In what follows we shall be interested in the spectral properties of 
the Hamiltonian
near the mobility edge $E_m$. 
There are several specific features intrinsic for this regime
supporting the multifractal behavior.
\begin{itemize}
\item

First, the level spacing distribution $P(s)$ is the key indicator
of the localization/delocalization transition. It behaves as 
\beq
\left\{
\begin{array}{ll}
P_{deloc}(s)= A\, s\,e^{-Bs^2} & \mbox{delocalized phase(GOE)}
\medskip \\ P_{loc}(s) = e^{-\frac{s}{2\chi}} & \mbox{localized phase}
\end{array} \right.
\label{eq:05}
\eeq
The parameter $\chi$ in the Poisson tail is  the level compressibility defined as 
\beq
\chi = \frac{d}{d\bar{n}}<(n-\bar{n})^2>, \qquad N\gg \bar{n}\gg 1
\eeq

\item
The second indicator is the  spectral formfactor which develops a fractal behavior at the mobility
edge which differs both from Wigner-Dyson and Poisson statistics \cite{edge}
\beq
R(\lambda)
\propto \lambda ^{-1+\frac{D_2}{d}}
\eeq
where $D_2$ is the fractal dimension defined as 
\beq
\sum_{r,n}<|\Psi_n(r)|^{2p}\delta(E-E_n)>\propto L^{-D_p(p-1)}
\eeq
in the volume $L^d$, d is dimension of space. 
The level number variance 
behaves  as  $\Sigma_{crit} \propto \chi E$ 
in the multifractal case, where the level compressibility reads as
\beq
\chi =  \frac{d-D_2}{2d}
\eeq

\item
There is also specific
violation of the normalization sum rules which yields the proper
behavior for the level number variance
\beq
(1-\chi) = 2 \int_{s>0}^{\infty} (1 -R_{\infty}(s'-s))ds'
\eeq
The expression above is written for $\beta=2$ , where $\beta$ is parameter 
of the $\beta$-ensemble ,  in the limit $e^{\frac{1}{\chi}}\gg 1$.

\end{itemize}

\subsection{Matrix models for localization transition}

Turn now for the specific matrix models used for the 
critical statistics and multifractality of eigenstates. There are several critical matrix 
models \cite{crit1,crit2,crit3} describing
the  localization transition in 3D 
which are qualitatively unified in \cite{muttalib}. All of them 
works only nearby the mobility edge
(see \cite{kravtsovrev} for review). Let us summarize their main features 

\begin{itemize}
\item
The two-matrix model \cite{crit1} with the
following probability function 
\beq
P(H) \propto \exp(-\beta TrH^2 - \beta b Tr([\Omega,H][\Omega,H]^{\dagger})
\eeq
where b is parameter and $\Omega$ is the fixed unitary matrix $\Omega = diag(\exp(\frac{2\pi i k}{N}))$.
The critical regime in this model implies that unitary symmetry breaking parameter 
behaves as $b=\mu N^2$ at $N\rightarrow \infty$. The matrix models of such types
appeared in the context of the non-singlet sectors in $c=1$ string theory \cite{klebanov,boulatov,maldacenanon}. However
contrary to that case we have no integration over $\Omega$ here and one could imagine
that a kind of saddle-point value is taken into account.

\item

The second  one-matrix model  involves the potential providing
the weak confinement \cite{crit2} of eigenvalues. Such type of model was used in the 
matrix model description of Chern-Simons theory \cite{tierz}
and was solved in terms of the q-Hermite polynomials. 
Asymptotically potential behaves as 
\beq
V(x)\rightarrow c\log^2 x, \qquad x\rightarrow \infty
\eeq
where $c$ - is some parameter and the probability measure
in the matrix integral reads as 
\beq
P(H)\propto \exp(-\beta TrV(H))
\eeq
The similar critical model for the  chiral ensembles has been 
considered in \cite{vercrit}.

\item

The third model was suggested in \cite{crit3} and involves the Gaussian
ensemble with independent random entries $(i\geq j)$
\beq
<H_{ij}>=0,\qquad <(H_{ij})^2>=\beta^{-1}[1 +\frac{(i-j)^2}{B^2}]^{-1}
\eeq
where B is parameter of the model, for $B>>1$ it gets mapped into supersymmetric sigma model. This 
model also manifests the multifractal behavior \cite{crit3} at the mobility edge

\end{itemize}

The  spectral correlators for all three
critical models are the same
\beq
R(E,s)=<\rho(E)\rho(E+s)>= \delta(s) + Y(E,s)
\eeq
where at small s
\beq
Y(E,s)\propto \frac{\pi^2\eta^2}{4} \frac{sin^2(\pi s)}{sinh^2(\pi^2 s \eta/2)}
\qquad \beta=2
\eeq
The parameter $\eta$ is related with the parameters of the matrix ensembles
$\eta=\frac{c}{\pi^2}= \mu $ if we assume $\eta\ll 1$. The spectral correlator 
in this regime is identical to 
the density-density correlator for the free fermion gas at finite temperature
proportional to $\eta$. The parameter $\eta $ is related with the fractal
dimension as $\eta = 1 - D_2$. Similarly at small $\eta$ regime 
the spectral compressibility reads as
\beq
\chi=1 + \int_{-\infty}^{+\infty} Y(E,s)ds
\eeq
and is consistent with the general relation
\beq.
\chi=\frac{d-D_2}{2d}
\eeq

Let us remark that formulas above are valid only for small
multifractality. At large s there is the power-tail
which knows about the fractal dimension as well.
According to our conjecture the small s regime
corresponds the IR region near the horizon while
the large s regime captures the information about
the UV scale. Since we are dealing with a kind of anomaly
phenomena the information about the fractal dimension
can be captured both in UV and IR regions.

\section{Mobility edge and BH horizon. Field theory}

Let us present a few qualitative arguments supporting the identification
of the near horizon region in the holographic dual and the 
mobility edge in the Dirac operator spectrum. 
\begin{itemize}
\item
At the critical metal-insulator transition in the Dirac operator spectrum
one could expect the jump of the chiral conductivity in the thermal QCD.
The corresponding Kubo-like formula for the correlator of the 
Noether currents generated left and right chiral rotations reads as 
\beq
i\int dx <J^{L}_{\nu}(x)J^{R}_{\mu}(0)>  = -\frac{1}{4}\eta_{\mu\nu} F_{\pi}^2
\eeq
Comparison of this QCD low-energy theorem with the 
formulas known in the transport phenomena provides the identification of the $F_{\pi}$ as the
diffusion coefficient in the chiral matter \cite{zahed}. Hence we could ask when
the jump of chiral conductivity is expected in the holographic 
setup. To be as model independent as possible consider the anomaly matching 
Son-Yamamoto condition in holographic QCD \cite{son} which yields the relation 
between the 2- and 3-point functions 
and is diagonal with respect to the holographic RG flows \cite{dgm}.
The Son-Yamamoto relation amounts to the following expression for the "`running"'$F_{\pi}$
\beq
F_{\pi}^{-2}(z) = \int_0^{z} \frac{1}{f^2(z)}
\eeq
which is valid for any reasonable holographic metric.
Taking the derivative of this expression we see immediately that
the criticality for the conductivity takes place exactly at the BH
horizon when $f(z)=0$.

\item

The lattice QCD studies demonstrate that the positions of mobility edge $\lambda_m(T)$
at $T>T_c$ grows as the function of the temperature near the deconfinement 
phase transition approximately as $\sqrt{T-T_c}$ \cite{kovacs}. Let us argue that this 
behavior  qualitatively agrees with the holographic picture upon the proper 
identification of the $\lambda=0$ point in the radial coordinate in the BH geometry. To 
justify this identification it is suitable 
to  use the Casher-Banks relation which relates the spectral density at the origin
with the chiral condensate. The chiral condensate is identified
with the tachyonic mode in the open string connecting the left and right D8 branes
near the tip of the cigar \cite{sonnen}. Hence $\lambda=0$ is naturally identified with the
position of the tip of the cigar.  Nearby the transition point 
we could inspect the relation between the horizon scale and the temperature
for AdS BH
\beq
f(r_h)=0 \qquad f(r)= 1+r^2 - \frac{M_{BH}}{r^2}\qquad \beta = \frac{4\pi}{f'(r_h)}
\eeq
hence near the maximum of the function $T^{-1}(r_h)$ the
relation holds
\beq
(T-T_c) \propto (r-r_h)^2
\eeq
This behavior is qualitatively consistent with 
the observed temperature dependence of the mobility edge if we assume that
$\lambda_m(T) \propto (r-r_h)$. 

Let us emphasize that the precise holographic metric in the deconfined
QCD is unknown however the presence of horizon is well established. The
detailed lattice study of the dependence $\lambda_m(T)$ could provide some
information concerning the holographic metric.  

\item

Recently the interesting relation for the spectral correlator has been 
found \cite{kanazawa}. The analogue of the Casher-Banks relation
for the spectral formfactor reads as follows
\beq
R(\lambda=0,\lambda=0) -R_{Pois}(\lambda=0,\lambda=0)= f_A
\label{kana}
\eeq
where $R(\lambda_1,\lambda_2)$ -is the spectral correlator of the Dirac 
operator in the deconfined phase and $f_A$ is defined as the coefficient 
in front of $U(1)_A$ symmetry breaking term
in the expansion of the partition function in terms of the fermion mass matrix M
\beq
Z(M)=\exp(-\frac{V_3}{T}(f_0 - f_2 Tr M^{+}M - f_A(det M + det M^{+}) +O(M^4)))
\eeq
The (\ref{kana})
measures the difference between the  spectral correlator in QCD
and spectral correlator in the case of Poisson statistics. We know
from the lattice studies that near the $\lambda=0$ the statistics
is Poisson hence this observation implies $f_A=0$ and unbroken $U(1)_A$ symmetry.
This issue is quite controversial and there are lattice results contradicting
and supporting this statement in the literature. This argument certainly
deserves for the additional study.

\end{itemize}

\section{Mobility edge and BH horizon. Worldsheet perspective}

The quark in the holographic representation corresponds to the
open string connecting the D8 brane placed at $m_q$ scale and
the IR region. 
In the confinement phase the string  is extended
along the holographic coordinate to the IR boundary and the tension of the
string remains finite at all values of the radial coordinate. 
In the deconfined phase the string is embedded in the target differently.
If a eigenvalue of the Dirac operator is large enough the main part of  open  
string is extended 
along the radial coordinate towards the horizon while the second part
of  string lies at the near-horizon region where it is almost 
tensionless. These two parts of  open string 
according to our conjecture correspond to the delocalized and localized eigenvalues of the 
Dirac operator correspondingly.

The arguments in the worldsheet theory go as follows.
Since a eigenvalue of the Dirac operator corresponds to the imaginary
quark mass the spectral properties of the quark wave function
in the boundary theory get mapped
into the  worldsheet theory on the open string extended in the radial coordinate. 
For  static quark the worldsheet of the open string involves the radial
coordinate and the Euclidean time $t_E=\tau$.
How 
the spectral statistics in the boundary theory gets translated into the worldsheet framework?
As we know the energy levels are independent in the localized
phase and interact repulsively in the delocalized phase.
The energy levels of the Dirac operator roughly correspond to  bits 
of the open string.  Using the 
spectrum - worldsheet correspondence we should analyze if the 
neighbor  "` string bits" are correlated or independent. In the confined
phase the string has the finite tension at all
distances therefore  tensionful string does not have any reason for 
uncorrelated neighbor bits. It is consistent with the
Wigner-Dyson  level statistics observed in lattice QCD in confined phase. In the 
deconfined phase the part of the string near horizon is almost 
tensionless and we conjecture that just this part of the string 
enjoys the Poisson statistics of the string bits.

To discuss the criticality from  the worldsheet theory viewpoint note that
the spectral correlator in the boundary theory gets mapped into correlator 
of densities of eigenvalues in the matrix model description
of  the string worldsheet theory
\beq
<\rho(E)\rho(E')>_{RMT}\rightarrow <\rho(x)\rho(x')>_{worldsheet}
\eeq
in the thermodynamical limit. Similar mapping is known for a while \cite{simons}
however in the holographic setting it acquires the very clear  origin
since the energy scale gets mapped into the value of the radial coordinate.
Moreover the bulk metric induces
the metric on the worldsheet, in particular in the deconfined 
phase the BH horizon induces the horizon on the worldsheet. The 
temperatures of the bulk and worldsheet black holes coincide for the 
static quark which has been clarified in the context of the 
holographic interpretation of drug force \cite{gubser}.

To get more insight on the interpretation of the density operator $\rho(x)$
in the worldsheet theory remind that the Wigner-Dyson RMT is equivalent to the 
ground state of the Calogero model for the fermions in the harmonic 
confining potential \cite{simons} with the Hamiltonian 
\beq
H_{Cal} =  \sum_{i=1}^{N} p_i^2 + \beta/2(\beta/2 -1) \sum_{i<j}\frac{1}{(x_i - x_j)^2} +\omega^2 \sum_{i=1}^{N}x_i^2
\eeq
The collective field $\rho(x,\tau)$ for the Calogero model
is described in terms of the two-dimensional scalar field as follows
\beq
\rho(x,\tau)= \rho_0 + A\cos (2\pi x + 2\Phi(x,\tau))
\eeq 
Hence the density-density correlator can be expressed in terms of the conventional 
Green function of the scalar $\Phi(x,\tau)$ in $(1+1)$. This is the way how 
the boundary spectral correlator in the confined phase is reproduced
in the worldsheet theory.

It turns out \cite{ggverba2003} that the RMT-Calogero correspondence at zero temperature
gets generalized to the relation between
the Calogero model at finite temperature and the critical
matrix model \cite{crit2}. Once again the
spectral correlator in the critical RMT gets mapped into the density-density correlator
in the Calogero model at finite temperature upon the identification deformation parameter b
in the matrix model as  \cite{ggverba2003}
\beq
2b=\frac{\omega}{sinh(\frac{\omega}{T})} \qquad 2b+1=\frac{\omega cosh(\frac{\omega}{T}) }{sinh(\frac{\omega}{T})} 
\eeq
Let us remind that coordinates of particles $x_i$ in Calogero model are their radial holographic coordinates while
$\tau$ is the Euclidean time identified with the angular coordinate at the hyperbolic plane.
The density-density correlator is taken at the same Euclidean time. The Calogero model is 
considered at the fixed coupling constant which corresponds to the fermions.

Note that we can also  use the duality found in \cite{nekrasov} to map the rational
Calogero model with the confining potential to the Sutherland model with the
trigonometric interaction using the polar decomposition
\beq
H_{Suth} = \sum_{i=1}^{N} p_i^2 + \beta /2 (\beta /2 -1) \sum_{i<j}\frac{1}{sin^2(x_i - x_j)}
\eeq
The time coordinate in Calogero model gets mapped into the space coordinate in the Sutherland model.
Both have interpretation as the Euclidean time in 4D Euclidean gauge theory.
From the viewpoint of the Calogero model the radius of the Euclidean time coordinate 
corresponds to the inverse temperature while in the Sutherland model it yields the
radius of the circle where the particles live. The density-density correlator 
in the Calogero model gets mapped into more complicated "`mixed"' correlator in the
trigonometric Sutherland model.  

The Sutherland representation of spectral correlator provides one more link with the
disorder picture. It was found in \cite{cardy} that the Sutherland model at the
boundary of the hyperbolic plane has the interpretation in terms the multiple interacting
radial SLE stochastic process. The strings are extended from the boundary into the bulk
and are affected by the random force.
If  string ends in the bulk obey the SLE dynamics the boundary degrees of freedom 
evolve under the Sutherland Hamiltonian. In our worldsheet picture it means
that we have large number of strings extended into radial direction  in the
random environment which are attached to the vortices on the string worldsheet.

Another way to get the same 
critical behavior is to consider the scalar field not on the 
thermal cylinder but in the 2d curved background. It was observed 
in \cite{kravtsovbh} that the background can be curiously identified
with the effective 2d "`acoustic BH"'. In fact the only
important point we need is that the background provides the 
horizon and effective Hawking temperature. This viewpoint 
reminds logic of  our study when the 2d BH metrics 
on the worldsheet is induced
from the bulk involving BH. The fractal dimension $D_2$ derived in \cite{kravtsovbh} is related 
to the Hawking  effective temperature T as
\beq
T\propto\pi \frac{d-D_2}{d}
\eeq

To get more insight on the interpretation of collective field $\Phi(s,\tau)$ 
in the worldsheet theory remind
that the $\cos\Phi$ terms appeared in the description of the string
in the 2d BH background \cite{kostov}. They came from the modes
with the unit winding around the compact dimension. It was argued \cite{kostov}
that the condensation of winding modes takes place near the BH horizon. 
Hence we could suggest that in our study the scalar field has the
same meaning as the collective field for the vortices. 
The vortices come from the non-singlet
sector in the $c=1$ matrix model \cite{klebanov,boulatov,maldacenanon}.
It is these vortices which yield the Calogero model and the Luttinger
effective collective field description.

Inspiring remark concerns the semiclassical expression for the spectral formfactor $K(t)$
in terms of the return probability $p(t)$ \cite{imry} in the non-relativistic limit
\beq
K(t)\propto |t|  p(t)
\label{imry}
\eeq
When $t$ is the usual time the return probability has the evident interpretation
in terms of the space-time trajectories. In that case (\ref{imry})
relates the transport properties connected with the return probability 
and the spectral properties of the system.  However in our study the "`time"' variable
has the meaning of the RG scale hence the formula like (\ref{imry}) has to be reinterpreted.
Naively the return probability in the RG flows has the meaning of the cyclic RG 
flow which occurs in several examples involving two charge RG flows. It
would be very interesting to recognize if the cyclic RG flows are important
for the  spectral formfactor in the worldsheet description. Remark that in our study 
we deal with the relativistic fermion in four dimensions hence (\ref{imry}) has to be 
generalized.

It is worth to mention the relation of our long string arguments
with two other problems. First, it was suggested in \cite{susskind}
that the deconfinement phase transition is related to the condensation
of the long strings forming the horizon.
Another related problem concerns the holographic interpretation of the 
entropic forces relevant for the meson dissociation \cite{kharzeev}
in the deconfined phase. The string connecting the quark and antiquark has 
a fragment lying at the horizon radial coordinate and it was argued that just this
fragment provides the large coordinate dependent entropy and 
therefore the entropic force. It looks  that the fragment
at the horizon radial scale is effectively broken into the independent bits. This is somewhat
parallel to our case  hence we could claim that the string bits at the horizon
in the dissociation problem  enjoy the Poisson statistics. The potential 
mechanism which would relate the Poisson statistics and the dissociation
phenomena is the formations of the multiple holes in the string worldsheet
in the near horizon limit. In the Euclidean regime it would yield
the independence of the neighbor string bits while upon the 
continuation to the Minkowski space it will corresponds to the decay of the
confining string into the multiple fragments since the Euclidean worldsheet
with holes serves as the bounce configuration for the string decay process.

\section{Conclusion}

Our study provides some evidences that the mobility edge in the Euclidean 4D Dirac
operator spectrum  in the deconfined phase of QCD corresponds to the BH near horizon region
in the holographic dual however further clarification
is certainly required. 
It is natural to assume that this correspondence is
quite general phenomenon and in particular the 2d BH in the dilaton JT gravity  
could be dual  to SYK model \cite{syk}(see \cite{rosenhaus} for review and references) only 
if we deform it to provide the criticality
in its spectrum. The precise example of the emerging 
mobility edge in the perturbed SYK has been found recently in \cite {garcia}.  Hence it is natural to conjecture
that the mobility edge in spectrum of Hamiltonian of the properly perturbed SYK model corresponds
to the 2D BH near horizon region. 

It would be also interesting to match our study with the discussion in \cite{Lee}. It was argued 
there that  quite generically criticality appears in the ungapped phase of boundary Euclidean theory
at the holographic horizon and its origin is the lack of possibility to match the initial UV state
with the particular state in IR  through the holographic RG flow. 
The tensor network language has been used there
to clarify this statement. Our study suggests that the corresponding criticality
at horizon is expected to be the metal-insulator type transition 
in the spectrum of the boundary Euclidean theory. 

Another important point suggested by our study is that it would be very 
interesting to investigate in detail in the lattice QCD framework the low-energy eigenmodes
of the Dirac operator in the deconfined QCD. It would provide the 
important information concerning the near-horizon region of the 
BH in 5D. The very confinement phenomena corresponds to the 
decay of the BH horizon into thermal gravitons. It would be also
interesting to investigate the case of deconfinement transition
induced by baryonic density from the viewpoint of Dirac operator spectrum.

In \cite{knots} the interesting relation between the condensate and 
torus knot invariants has been found. It turned out that 
a fermion condensate in $SU(2)$ SUSY QCD plays the role of generating 
function for the torus knot invariants. Torus knots are placed in the 
internal "`momentum"'  space however they are intimately related with the invariants of the
moduli space of instantons sitting at one point on the top of each other
in the physical Euclidean 4D space-time. It would
be interesting to question if similar relation between the torus knot invariants
and  the Dirac operator spectrum exists in the instanton-antiinstanton 
vacuum ensemble in QCD.

 I am grateful to M.Litvinov and N.Sopenko for collaboration  and to V. Braguta, A. Kamenev, D. Kharzeev, 
V. Kravtsov, A. Milekhin, S. Nechaev, N. Prokof'ev , B. Shklovskii  for the useful discussions. 
The work was supported by Basis Foundation Fellowship.

\end{document}